\documentclass[a4paper]{jpconf}
\usepackage{graphicx}

\usepackage{amsmath, amsfonts, amssymb}
\usepackage{mathrsfs}

\usepackage{graphicx,epsfig}
\usepackage{epic}
\usepackage{color}

\newcommand{\adscp}{$AdS_{4}\times \mathbb{CP}^{3}$}
\newcommand{\ads}{$AdS_5\times S^5$}

\begin{document}
\title{Semiclassical energy of the $AdS_4 \times \mathbb{CP}^3$ folded string}

\author{Matteo Beccaria$^{(1,a)}$, Guido Macorini$^{(2,b)}$, CarloAlberto Ratti$^{(1,c)}$ and Saulius Valatka$^{(3,d)}$} 
\address{$^1$\ Dipartimento di Matematica e Fisica ``Ennio De Giorgi``, Universita' del Salento \& INFN, Via Arnesano, 73100 Lecce, Italy} 
\address{$^2$\ Niels Bohr International Academy and Discovery Center, Niels Bohr institute,
Blegdamsvej 17 DK-2100 Copenhagen, Denmark}
\address{$^3$\ Mathematics Department, Kings College London, The Strand, London WC2R 2LS, UK}

\ead{$^a$\ matteo.beccaria@le.infn.it,\hspace{5mm} $^b$\ macorini@nbi.ku.dk,\\ \hspace{10mm} $^c$\ carloalberto.ratti@le.infn.it, $^d$\ saulius.valatka@kcl.ac.uk}

\begin{abstract}
We consider the classical solution describing a folded type IIA string in the background $AdS_4 \times \mathbb{CP}^3$. The string is spinning in $AdS$ and has angular momentum in $\mathbb{CP}^3$. In the 't Hooft limit, this is the gravity dual of twist operators in the ABJM superconformal theory. We quantize the classical solution by algebraic curve methods and determine the first semiclassical correction to the energy. An integral representation is given, valid for all values of the charges. We analyze its properties in the special regimes associated with a short or long string. Finally, we investigate various properties of the leading term of the energy for short strings (the so-called slope).
\end{abstract}

\section{Introduction}

The AdS/CFT correspondence
is an extremely fruitful theoretical idea that played a central role in the past. 
In its simplest instance, it links the integrability of the world-sheet $\sigma$-model for type IIB superstring
to the strong coupling behaviour of four dimensional gauge theories. In the specific case of planar $\mathcal{N}=4$
Super Yang-Mills theory, integrability methods determine the spectrum of anomalous dimensions as a function of 
the 't Hooft coupling $\lambda$, from weak to strong coupling. More precisely, the spectrum was shown to be captured by suitable Y-system equations \cite{Gromov:2009tv,Arutyunov:2009zu,Bombardelli:2009ns,Gromov:2009bc,Arutyunov:2009ur}. These equations are expected to be general and valid for any operator, including short ones. The first attempt of explorations of the flow of the anomalous dimension concerned the simplest $\mathfrak{sl}(2)$ sector of the theory and the simplest operator present in this sector, the Konishi operator. An important step has been the  determination of the analytical strong coupling expansion of the Konishi anomalous dimension, {\em i.e.} the energy of its gravity dual state. The sub-leading coefficient in the string theory expansion was found in \cite{Gromov:2011de,Roiban:2011fe,Vallilo:2011fj}. A highly nontrivial observation which allows one to reproduce the one-loop result almost without efforts was made recently in \cite{Basso:2011rs}. The second nontrivial strong coupling expansion coefficient was derived analytically in \cite{Gromov:2011bz} by using the one-loop expression for a general  $(S,J)$ folded string \cite{Frolov:2002av}.

The folded string is indeed a very useful theoretical laboratory. At the classical level, it can be described in terms of the scaled charges $\mathcal{S} = S/\sqrt\lambda$, $\mathcal{J} = J/\sqrt\lambda$. The spin $\mathcal{S}$ is associated with rotation around the center of  $AdS_{5}$ while $\mathcal{J}$ is an angular momentum on the sphere. At small $\mathcal{S}$, the string is short and admits
a near-flat space description. At large $\mathcal{S}$, the string stretches and reaches the boundary of $AdS$ with 
the characteristic scaling $E \sim \log \mathcal{S}$ of its energy. Quantum corrections in the semiclassical approximation
are fully under control and can be studied by algebraic curve tools.

Remarkably, very similar integrability structures as well as folded string classical solutions are known to be 
available in the ABJM theory~\cite{Aharony:2008ug}.
A comprehensive review of integrability methods in ABJM can be found in \cite{Klose:2010ki}. For our purposes it is important to remind that also ABJM presents a $\mathfrak{sl}(2)$-like sector.
At strong coupling and large spin these operators behave quite similarly to the corresponding ones in $AdS_5 \times S^5$. In particular their dual string state is a folded string 
rotating in $AdS_3$  and with angular momentum $J$ in $\mathbb{CP}^3$~\cite{McLoughlin:2008ms,Alday:2008ut,Krishnan:2008zs}. 
At weak coupling, they are composite operators in totally different theories.
The quantized folded string on $AdS_4 \times \mathbb{CP}^3$ have been studied in various limits. The large spin limit has 
been studied at leading order in two regimes: At $J/S$ fixed \cite{McLoughlin:2008ms} and at $J\sim \log S$ \cite{Gromov:2008fy}. 
The short string limit has never been addressed. On the numerical side, the analysis of \cite{LevkovichMaslyuk:2011ty} explored the Thermodynamic Bethe Ansatz (TBA) equations of $AdS_4 \times \mathbb{CP}^3$, although in a small window of the coupling reaching at most $\lambda \sim \mathcal{O}(1)$.
In paper \cite{folded}, a complete analysis of the one-loop semiclassical energy of folded string in $AdS_4 \times \mathbb{CP}^3$  has been presented. Here we review the main results found there.

The plan of the paper is the following: in Section \ref{AlgCurve} we summarize the general features of the quantization through algebraic curve in $AdS_4 \times \mathbb{CP}^3$. In Section \ref{Folded} we describe the specific features of the algebraic curve associated with the folded string solution. In Section \ref{4} we explain how to  compute the semi-classical energy of such solutions. The short string limits and the slope functions are analyzed respectively in Sections \ref{5} and \ref{sec:slope} while the long string limit is computed in Section \ref{7}. Our final prediction for short states in ABJM can be found in Section \ref{8}.

\section{Algebraic curve quantization in \adscp}
\label{AlgCurve}

In this section we give a compact self-contained summary of the results of \cite{Gromov:2008bz} using the 
language of off-shell fluctuation energies \cite{Gromov:2008ec}. We shall work in the algebraic curve 
regularization \cite{Gromov:2008fy} and write all equations in terms of the $\sigma$-model coupling $g$. For large $g$, it is 
related to the 't Hooft coupling by $\lambda = N/k = 8\,g^{2}$,
but, contrary to the \ads\ case, this relation will get corrections at finite $g$.
The classical algebraic curve for \adscp\ is a 10-sheeted Riemann surface. The spectral parameter moves on it and 
we shall consider 10 symmetric quasi momenta $q_{i}(x)$ 
\begin{equation}
(q_{1}, q_{2}, q_{3}, q_{4}, q_{5}) = (-q_{10}, -q_{9}, -q_{8}, -q_{7}, -q_{6}).
\end{equation}
They can have branch cuts connecting the sheets with $q_{i}^{+}-q_{j}^{-} = 2\,\pi\,n_{ij}$.
In the terminology of \cite{Gromov:2008bz}, the physical polarizations $(ij)$ can be split into {\em heavy} and {\em light} ones and are summarized in the following table:
 $$
 \begin{array}{c|ccc}
  & \mbox{AdS${}_{4}$} & \mbox{Fermions} & \mathbb{CP}^{3} \\
  \hline
  \mbox{heavy} & \quad (1,10) (2,9) (1,9)\quad & (1,7) (1,8) (2,7) (2,8) & (3,7) \\
  \mbox{light} & & (1,5) (1,6) (2,5) (2,6) & \quad (3,5) (3,6) (4,5) (4,6)\quad
  \end{array}
 $$
Inversion symmetry imposes these other constraints on the quasi momenta
\begin{equation}
q_{1}(x) = -q_{2}(1/x), \qquad
q_{3}(x) = 2\,\pi\,m-q_{4}(1/x), \qquad
q_{5}(x) = q_{5}(1/x),
\end{equation}
where $m\in\mathbb Z$ is a winding number. Thus, only 3 out of 10 quasi momenta are independent.

\subsection{Semiclassical quantization}

Semiclassical quantization is achieved by perturbing quasi-momenta introducing extra poles that shift the quasi-momenta
$q_{i}\to q_{i}+\delta q_{i}$.
The asymptotic expression of $\delta q_{i}$ are related to the number $N_{ij}$ of extra fluctuations and to the energy correction $\delta E$.
The off-shell frequencies $\Omega^{ij}(x)$ are defined in order to have 
\begin{equation}
\delta E = \sum_{n, ij} N^{ij}_{n}\,\Omega^{ij}(x^{ij}_{n}),
\end{equation}
where the sum is over all pairs $(ij)\equiv (ji)$ of physical polarizations and integer values of $n$ with $q_{i}(x^{ij}_{n})-q_{j}(x^{ij}_{n}) = 2\,\pi\,n$.
By linear combination of frequencies and inversion (as in the ${\rm AdS}_{5}/{\rm CFT}_{4}$ case), we can derive all  the off-shell frequencies in terms of two fundamental ones $\Omega_{A}(x) = \Omega^{15}(x)$, $\Omega_{B}(x) = \Omega^{45}(x)$.
Their explicit expressions turn out to be 
\begin{eqnarray}
&& \Omega^{29} =  2\,\left[-\Omega_{A}(1/x)+\Omega_{A}(0)\right], \nonumber \qquad \quad\ \ 
\Omega^{1, 10} =  2\,\Omega_{A}(x),\nonumber \\
&& \Omega^{19} =  \Omega_{A}(x)-\Omega_{A}(1/x)+\Omega_{A}(0), \nonumber \qquad \,
\Omega^{37} =\Omega_{B}(x)-\Omega_{B}(1/x)+\Omega_{B}(0), \nonumber \\
&& \Omega^{35}  = \Omega^{36} = -\Omega_{B}(1/x)+\Omega_{B}(0),\nonumber \qquad \ 
\Omega^{45} = \Omega^{46} = \Omega_{B}(x), \nonumber \\
&&\Omega^{17} = \Omega_{A}(x)+\Omega_{B}(x), \nonumber \qquad \qquad \qquad \ 
\Omega^{18} = \Omega_{A}(x)-\Omega_{B}(1/x)+\Omega_{B}(0), \nonumber \\
&& \Omega^{27} = \Omega_{B}(x)-\Omega_{A}(1/x)+\Omega_{A}(0), \nonumber \qquad
\Omega^{28} = -\Omega_{A}(1/x)+\Omega_{A}(0)-\Omega_{B}(1/x)+\Omega_{B}(0), \nonumber \\
&& \Omega^{15} = \Omega^{16} = \Omega_{A}(x), \nonumber \qquad \qquad \qquad \quad \ 
\Omega^{25} = \Omega^{26} = -\Omega_{A}(1/x)+\Omega_{A}(0).
\end{eqnarray}

\section{The folded string in $AdS_4 \times \mathbb{CP}^3$}
\label{Folded}

We present the algebraic curve for the folded string in \adscp\ closely following  the notation of  \cite{Gromov:2008fy}.
In terms of the semiclassical variables $\mathcal S = \frac{S}{4\,\pi\,g}$ and $\mathcal J = \frac{J}{4\,\pi\,g}$,
the energy of the folded string can be expanded according to 
\begin{equation}
E = 4\,\pi\,g\,\,\mathcal E_{0}(\mathcal J, \mathcal S)+E_{1}(\mathcal J, \mathcal S)+\mathcal O\left(\frac{1}{g}\right),
\end{equation}
where the small $\mathcal S$ expansion of the classical contribution $\mathcal E_{0}$ reads
\begin{equation}
\label{eq:classical}
 \mathcal E_{0}= \mathcal J+\frac{\sqrt{\mathcal J^{2}+1}}{\mathcal J}\,\mathcal S-\frac{\mathcal J^{2}+2}{4\,\mathcal J^{3}(\mathcal J^{2}+1)}\,\mathcal S^{2}+ 
 \frac{3\,\mathcal J^{6}+13\,\mathcal J^{4}+20\,\mathcal J^{2}+8}{16\,\mathcal J^{5}\,(\mathcal J^{2}+1)^{5/2}}\,\mathcal S^{3}+\cdots.
\end{equation}
The quasi momenta are closely related to those of the $AdS_5 \times S^5$ folded string since motion is still in $AdS_{3}\times S^{1}$
and the $\mathbb{CP}^{3}$ part of the background plays almost no role. The only non trivial case is 
\begin{eqnarray}
q_{1}(x) &=& \pi\,f(x)\,\left\{-\frac{J}{4\,\pi\,g}\,\left(\frac{1}{f(1)\,(1-x)}-\frac{1}{f(-1)(1+x)}\right) 
-\frac{4}{\pi\,(a+b)(a-x)(a+x)}\ \times \right. \nonumber \\
&& \left.\qquad \ \ \ \ \  \times \left[
(x-a)\,\mathbb K\left(\frac{(b-a)^{2}}{(b+a)^{2}}\right)
+ 2\,a\,\Pi\left(\left.
\frac{(b-a)(a+x)}{(a+b)(x-a)} \right| \frac{(b-a)^{2}}{(b+a)^{2}}
\right)
\right]
\right\}-\pi,
\end{eqnarray}
where $1<a<b$, $f(x) = \sqrt{x-a}\,\sqrt{x+a}\,\sqrt{x-b}\,\sqrt{x+b}$ and
\begin{eqnarray}
&& S = 2\,g\,\frac{ab+1}{ab}\,\left[b\,\mathbb E\left(1-\frac{a^{2}}{b^{2}}\right)
-a\,\mathbb K\left(1-\frac{a^{2}}{b^{2}}\right)\right], 
\quad \nonumber 
J = \frac{4\,g}{b}\,\sqrt{(a^{2}-1)(b^{2}-1)}\,\mathbb K\left(1-\frac{a^{2}}{b^{2}}\right), \\
&&\phantom{MBnmnpfdmdqctpdm} E = 2\,g\,\frac{ab-1}{ab}\,\left[b\,\mathbb E\left(1-\frac{a^{2}}{b^{2}}\right)
+a\,\mathbb K\left(1-\frac{a^{2}}{b^{2}}\right)\right].
\end{eqnarray}
The other independent quasi-momenta are $q_{3}(x) = \frac{J}{2\,g}\,\frac{x}{x^{2}-1}$, $q_{5}(x) = 0$.
The above expressions are valid for a folded string with minimal winding. Adding winding is trivial at the classical level, but requires non trivial changes at the one-loop level \cite{Gromov:2011bz}.

The independent off-shell frequencies can be determined by the methods of \cite{Gromov:2008ec} and read
\begin{eqnarray}
&& \Omega_{A}(x) =  \frac{1}{ab-1}\left(1-\frac{f(x)}{x^{2}-1}\right),   \qquad \Omega_{B}(x) = \frac{\sqrt{a^{2}-1}\,\sqrt{b^{2}-1}}{ab-1}\,\frac{1}{x^{2}-1}.
\end{eqnarray}

\section{Integral representation for the one-loop correction to the energy}
\label{4}

The one-loop shift of the energy is given in full generality by the following sum of zero point energies
\begin{equation}
\label{eq:one-loop-correction}
E_{1} = \frac{1}{2}\,\sum_{n=-\infty}^{\infty}\,\sum_{ij}(-1)^{F_{ij}}\,\omega^{ij}_{n},\qquad
\omega_{n}^{ij} = \Omega^{ij}(x^{ij}_{n}),
\end{equation}
where the sum over $ij$ is over the $8_{B}+8_{F}$ physical polarizations and $x^{ij}_{n}$ is the unique solution 
to the equation $q_{i}(x^{ij}_{n})-q_{j}(x^{ij}_{n}) = 2\,\pi\,n$ under the condition
$|x_{n}^{ij}|>1$~\footnote{ If it happens that for some $ij$ and $n$ the above equation has no solution, then we shall say that the polarization $(ij)$ has the { missing mode} $n$. Missing modes can be treated according to the procedure
discussed in \cite{Gromov:2008ec}.
}.
The explicit sum over the infinite number of on-shell frequencies requires some care and a definite prescription 
since the sums are not absolutely convergent due to cancellations between bosonic and
fermionic contributions (see \cite{folded} for an exhaustive discussion about this point).

In an alternative approach \cite{Gromov:2011de,Gromov:2011bz}, the infinite sum (\ref{eq:one-loop-correction})
can be evaluated by contour integration in the complex plane. The result is quite similar to the \ads\ one and reads
\begin{equation}
E_{1} = E_{1}^{\rm anomaly, 1}+ 2 E_{1}^{\rm anomaly, 2}+E_{1}^{\rm dressing}+E_{1}^{\rm wrapping},
\end{equation}
where, by defining $x(z) = z+\sqrt{z^{2}-1}$, we have 
\begin{eqnarray}
E_{1}^{\rm anomaly, 1} &=& 2\,\int_{a}^{b}\,\frac{dx}{2\,\pi\,i}\left[\Omega^{1,10}(x)-\Omega^{1,10}(a)\right]\,
\partial_{x}\,\log\sin q_{1}(x), \\
E_{1}^{\rm anomaly, 2} &=& 2\times 2\,\int_{a}^{b}\,\frac{dx}{2\,\pi\,i}\left[\Omega^{1,5}(x)-\Omega^{1,5}(a)\right]\,\partial_{x}\,\log\sin \frac{q_{1}(x)}{2}, \\
E_{1}^{\rm dressing} &=& \sum_{ij}(-1)^{F_{ij}}\,\int_{-1}^{1}\frac{dz}{2\,\pi\,i}
\,\Omega^{ij}(z)\,\partial_{z}\frac{i\,\left[q_{i}(z)-q_{j}(z)\right]}{2},\\
E_{1}^{\rm wrapping} &=& \sum_{ij}(-1)^{F_{ij}}\,\int_{-1}^{1}\frac{dz}{2\,\pi\,i}
\,\Omega^{ij}(z)\,\partial_{z}\log(1-e^{-i\,(q_{i}(z)-q_{j}(z))}),
\end{eqnarray} 
As in \ads, the labeling of the various contributions reminds their physical origin. In particular, dressing and wrapping contributions are separated so that the asymptotic contribution is divided from finite size effects. As in \ads, the anomaly terms are special contributions arising from the deformation of contours and ultimately due to the presence of the algebraic curve cuts.
The representation (\ref{eq:one-loop-correction}) is a compact formula for $E_{1}$ and can be evaluated numerically with minor effort. We shall now analyze the short and long string limit. 

\section{Short string limit}
\label{5}

The short string limit is generically $\mathcal S\to 0$. Regarding $\mathcal J$, we shall consider two cases. The first amounts to 
keeping $\mathcal J$ fixed, expanding in the end each coefficient of powers of $\mathcal S$ at small $\mathcal J$.
This is the procedure worked out in \cite{Gromov:2011bz} in \ads. In the second case, we shall keep the ratio
$\rho = \mathcal J/\sqrt\mathcal S$ fixed as in \cite{Beccaria:2012tu}. The two expansions are related, but not equivalent and provide useful different information.
\begin{itemize}  
\item[] {\bf Fixed $\mathcal J$ expansion} \\
\end{itemize}
\vspace{-4mm}
After a straightforward computation, our main result is 
\begin{eqnarray}
\label{eq:GV}
E_{1} &=& 
\bigg(
-\frac{1}{2 \mathcal{J}^2}+\frac{\log (2)-\frac{1}{2}}{\mathcal{J}}+\frac{1}{4}+\mathcal{J} \left(-\frac{3 \,\zeta (3)}{8}+\frac{1}{2}-\frac{\log (2)}{2}\right)-\frac{3
   \mathcal{J}^2}{16}+\\
   &&+\mathcal{J}^3 \left(\frac{3 \,\zeta (3)}{16}+\frac{45 \,\zeta (5)}{128}-\frac{1}{2}+\frac{3 \log (2)}{8}\right)+
   \cdots
\bigg)\,\mathcal S+ \nonumber \\
&& +\bigg(
\frac{3}{4 \mathcal{J}^4}+\frac{\frac{1}{2}-\log (2)}{\mathcal{J}^3}-\frac{1}{8 \mathcal{J}^2}+\frac{\frac{1}{16}-\frac{3 \,\zeta (3)}{4}}{\mathcal{J}}-\frac{1}{8}+\mathcal{J} \left(\frac{69 \,\zeta
   (3)}{64}+\frac{165 \,\zeta (5)}{128}-\frac{27}{32}+\frac{\log (2)}{2}\right)+\nonumber \\
   && +\frac{3 \mathcal{J}^2}{8}+\mathcal{J}^3 \left(-\frac{163 \,\zeta (3)}{128}-\frac{405 \,\zeta (5)}{256}-\frac{875 \,\zeta
   (7)}{512}+\frac{235}{128}-\log (2)\right)+\cdots
\bigg)\,\mathcal S^{2}+ \nonumber \\
&& + 
\bigg(
-\frac{5}{4 \mathcal{J}^6}+\frac{\frac{3 \log (2)}{2}-\frac{3}{4}}{\mathcal{J}^5}+\frac{\frac{9 \,\zeta (3)}{16}+\frac{1}{16}}{\mathcal{J}^3}+\frac{1}{16 \mathcal{J}^2}+\frac{\frac{45 \,\zeta
   (3)}{64}+\frac{75 \,\zeta (5)}{256}-\frac{7}{32}+\frac{\log (2)}{8}}{\mathcal{J}}+\frac{11}{64}+\nonumber \\
   &&  +\mathcal{J} \left(-\frac{89 \,\zeta (3)}{32}-\frac{745 \,\zeta (5)}{256}-\frac{3815 \,\zeta
   (7)}{2048}+2-\frac{33 \log (2)}{32}\right)-\frac{465 \mathcal{J}^2}{512}+\nonumber \\
   && + \mathcal{J}^3 \left(\frac{5833 \,\zeta (3)}{1024}+\frac{1585 \,\zeta (5)}{256}+\frac{98035 \,\zeta (7)}{16384}+\frac{259455
   \,\zeta (9)}{65536}-\frac{405}{64}+\frac{775 \log (2)}{256}\right)+\cdots
\bigg)\,\mathcal S^{3}+ \cdots\nonumber
\end{eqnarray}
This expansion is rather similar to the one derived in \cite{Gromov:2011bz} for \ads, but there are two remarkable differences:
1) The leading terms at small $\mathcal J$ are $\mathcal O(\mathcal S^{n}/\mathcal J^{2n})$. Instead, they were $\mathcal O(\mathcal S^{n}/\mathcal J^{2n-1})$ in \ads. Also, there are terms with all parities in $\mathcal J$ while in \ads, there appear only terms odd under $\mathcal J\to -\mathcal J$. The additional terms
imply that if one scales $\mathcal J\sim \sqrt\mathcal S$, there is a constant contribution in the short string limit. This is different 
from \ads\ where the energy correction vanishes like $\sqrt\mathcal S$ in this regime;
2) There are terms proportional to $\log(2)$. These terms can be removed by redefining $g\longrightarrow g-\frac{\log 2}{4\,\pi}$.
This redefinition connects the coupling in the algebraic curve regularization with the coupling in the world-sheet regularization and agrees with the calculation in \cite{McLoughlin:2008he} for the circular string solution as well as in \cite{Abbott:2010yb} for the giant magnon.

\begin{itemize}
\item[] {\bf Fixed $\rho = \mathcal J / \sqrt\mathcal S$ expansion} \\
\end{itemize}
\vspace{-6mm}
The result in this limit is 
\begin{eqnarray}
\label{eq:our-expansion}
&& E_{1} = -\frac{1}{2}\,\mathcal C(\rho, \mathcal S)+a_{01}(\rho)\,\sqrt\mathcal S+ a_{1,1}(\rho)\,\mathcal S^{3/2}+\mathcal O(\mathcal S^{5/2}),
\end{eqnarray}
where
\begin{eqnarray}
a_{1,0}(\rho) &=& \frac{2\,\log (2)-1}{2\,\sqrt{\rho^{2}+2}}, \\
a_{1,1}(\rho) &=& -\frac{\log (2)\left(2 \rho ^4+6 \rho ^2+3\right)}{4 \left(\rho ^2+2\right)^{3/2}}+\frac{8 \rho ^4+25 \rho ^2+16}{16 \left(\rho ^2+2\right)^{3/2}}-\frac{3 \left(\rho
   ^2+3\right) \zeta (3)}{8 \sqrt{\rho ^2+2}}.
\end{eqnarray}
$\mathcal C$ is related to the branch cut endpoints by the formula
\begin{equation}
\label{eq:theC}
\mathcal C = \frac{\sqrt{(a^{2}-1)\,(b^{2}-1)}}{1-a\,b}+1.
\end{equation}
Its expansion at small $\mathcal S$ with fixed $\rho=\mathcal J/\sqrt\mathcal S$ is 
\begin{equation}
\mathcal C = 1-\frac{\rho}{\sqrt{\rho^{2}+2}}-\frac{2\,\rho^{3}+5\,\rho}{4\,(\rho^{2}+2)^{3/2}}\,\mathcal S + 
\frac{\rho\,(12\,\rho^{6}+68\,\rho^{4}+126\,\rho^{2}+73)}{32\,(\rho^{2}+2)^{5/2}}\,\mathcal S^{2}+\cdots
\end{equation}
Notice also that upon expanding at small $\mathcal J$ the expansion of $\mathcal C$ at fixed $\mathcal J$ we get precisely the terms with even/odd $\mathcal J$ exponents in the coefficients of the odd/even powers of $\mathcal S$ in (\ref{eq:GV}).

Expanding $E_{1}$ at large $\rho$ we  partially resum the calculation at fixed $\mathcal J$. Just to give an 
example, from the expansion 
\begin{equation}
-\frac{1}{2}\left(1-\frac{\rho}{\sqrt{\rho^{2}+2}}\right) = -\frac{1}{2 \rho ^2}+\frac{3}{4 \rho ^4}-\frac{5}{4 \rho ^6}+\frac{35}{16 \rho
   ^8}-\frac{63}{16 \rho ^{10}}+\cdots,
\end{equation}
we read the coefficients of all terms $\sim \mathcal S^{n}/\mathcal J^{2n}$. The first ones are of course in agreement
with (\ref{eq:GV}). From the large $\rho$ expansion of $a_{11}(\rho)$ 
we can read the coefficients of all terms $\sim \mathcal S^{n} / \mathcal J^{2n-3}$ \cite{folded}.

\section{The slope function}
\label{sec:slope}

The one-loop correction $E_{1}$ tends to zero linearly with $\mathcal S$ when $\mathcal S\to 0$ at fixed $\mathcal J$.
The slope ratio
\begin{equation}
\sigma(\mathcal J) = \lim_{\mathcal S\to 0}\frac{E_{1}(\mathcal S, \mathcal J)}{\mathcal S},
\end{equation}
is an important quantity  related to the conjectures in \cite{Basso:2011rs}~\footnote{The exact slope mentioned in 
\cite{Basso:2011rs} is the coefficient of $S$ in the expansion of $E^{2}$. This line of analysis is suggested by the simplicity of the marginality condition in \ads (see \cite{Tseytlin:2003ac} for a general discussion). Here, it is simpler to discuss the quantity $\sigma(\mathcal J)$.} (see also \cite{BassoSlope,GromovSlope}). It is known that it does not receive dressing corrections both in \ads\ and in \adscp\ since such contributions start at 
order $\mathcal S^{2}$ \cite{Basso:2011rs}. It also 
does not receive wrapping corrections in $AdS_{5}$.  Instead, in the case of $AdS_{4}$ the slope has a non vanishing wrapping contribution. For instance, a rough evaluation at $\mathcal J=1$ gives  a definitely non zero value around $-0.042$.

Indeed, an analytical calculation shows that  the wrapping contribution to the slope in \adscp\ is exactly
\begin{equation}
\sigma^{\rm wrap}(\mathcal J) = \sum_{n=-\infty}^{\infty}\sigma_{n} = -\frac{1}{2\,\mathcal J}\,\sum_{n=-\infty}^{\infty}\frac{(-1)^{n}}{
\sqrt{\mathcal J^{4}+(n^{2}+1)\,\mathcal J^{2}+n^{2}}}.
\end{equation}
To analyze the small $\mathcal J$ limit it is convenient to split this contribution into the $n=0$ term plus the rest. The result is very intriguing. For the $n=0$ term, 
we find 
\begin{eqnarray}
\sigma^{\rm wrap}_{n=0} = -\frac{1}{2\,\mathcal J^{2}\,\sqrt{\mathcal J^{2}+1}} = 
-\frac{1}{2 \mathcal{J}^2}+\frac{1}{4}-\frac{3 \mathcal{J}^2}{16}+\frac{5 \mathcal{J}^4}{32}-\frac{35 \mathcal{J}^6}{256}+\frac{63 \mathcal{J}^8}{512}+\cdots\,.
\end{eqnarray}
This is precisely the set of terms even under $\mathcal J\to -\mathcal J$ in the full slope which is the first term of  (\ref{eq:GV}).
Similarly, we can consider the rest of $\sigma^{\rm wrap}$ and expand at small $\mathcal J$. We find 
\begin{eqnarray}
\sum_{n\neq 0}\sigma^{\rm wrap}_n &=& \frac{\log (2)}{\mathcal{J}}+\mathcal{J} \left(-\frac{3 \zeta(3)}{8}-\frac{\log (2)}{2}\right)+\mathcal{J}^3 \left(\frac{3 \zeta(3)}{16}+\frac{45 \zeta(5)}{128}+\frac{3 \log (2)}{8}\right)+\nonumber \\
&&+\mathcal{J}^5
   \left(-\frac{9 \zeta(3)}{64}-\frac{45 \zeta(5)}{256}-\frac{315 \zeta(7)}{1024}-\frac{5 \log (2)}{16}\right)+O\left(\mathcal{J}^6\right).
\end{eqnarray}
Comparing again with (\ref{eq:GV}), we see that we are reproducing all the irrational terms of the slope, involving zeta functions or $\log(2)$. The remaining terms 
are the same as in \ads
~\footnote{
This is due to the fact that the BAE are essentially the same as for $\mathfrak{sl}(2)$ sector in \ads.},
\begin{equation}
\sigma(\mathcal J) -\sigma^{\rm wrap}(\mathcal J) = -\frac{1}{2\,\mathcal J}+\frac{\mathcal J}{2}-\frac{\mathcal J^{3}}{2}+\cdots.
\end{equation}
Thus, we are led to the following expression for the one-loop full slope 
\begin{equation}
\sigma(\mathcal J) = -\frac{1}{2\,\mathcal J}\left[\frac{1}{\mathcal J^{2}+1}+\sum_{n=-\infty}^{\infty}\frac{(-1)^{n}}{
\sqrt{\mathcal J^{4}+(n^{2}+1)\,\mathcal J^{2}+n^{2}}}
\right].
\end{equation}
The above analysis of the slope is a confirmation  that the various terms in (\ref{eq:GV}) are organized in the expected way. The asymptotic contribution is precisely the same as in \ads, while wrapping is different and is exponentially suppressed for large operators \cite{folded}.

\section{Long string limit}
\label{7}

The large $S$ behaviour of the one-loop energy $E_{1}$ can be computed starting from the integral representation (\ref{eq:GV}).
Let us first summarize the result valid for \ads\ from \cite{Gromov:2011de}. We scale $\mathcal J$ with $\mathcal S$ for $\mathcal S\gg 1$ according to 
\begin{equation}
\mathcal J = \frac{\ell}{\pi}\,\log\left(\frac{8\,\pi\,\mathcal S}{\sqrt{\ell^{2}+1}}\right),
\end{equation}
where we assume $\ell>0$.
Then, the one-loop energy correction can be written
\begin{equation}
E_{1}^{AdS_{5}} = f_{10}^{AdS_{5}} (\ell)\,\log\left(\frac{8\,\pi\,\mathcal S}{\sqrt{\ell^{2}+1}}\right)+f_{11}^{AdS_{5}} (\ell)+\frac{c^{AdS_{5}}}{\log\left(\frac{8\,\pi\,\mathcal S}{\sqrt{\ell^{2}+1}}\right)}+\cdots ,
\end{equation}
with
\begin{eqnarray}
f_{10}^{AdS_{5}}(\ell) &=& \frac{\sqrt{\ell^2+1}+2 \left(\ell^2+1\right) \log
   \left(\frac{1}{\ell^2}+1\right)-\left(\ell^2+2\right)
   \log
   \left(\frac{\sqrt{\ell^2+2}}{\sqrt{\ell^2+1}-1}\right)-1
   }{\pi  \sqrt{\ell^2+1}}, \\
f_{11}^{AdS_{5}}(\ell) &=& \frac{2 \left(\log
   \left(1-\frac{1}{\left(\ell^2+1\right)^2}\right)+2
   \sqrt{\ell^2+1} \cot ^{-1}\left(\sqrt{\ell^2+1}\right)+2
   \coth ^{-1}\left(\sqrt{\ell^2+1}\right)-2 \ell \cot
   ^{-1}(\ell)\right)}{\pi  \sqrt{\ell^2+1}},\nonumber \\
c^{AdS_{5}}(\ell) &=& -\frac{\pi}{12\,(\ell^{2}+1)}.\nonumber
\end{eqnarray}

The expansion in \adscp\ can be derived in the same way as in \cite{Gromov:2011de} and the result is
\begin{equation}
\label{eq:AdS4-largeS}
E_{1}^{AdS_{4}} =f_{10}^{AdS_{4}} (\ell)\,\log\left(\frac{8\,\pi\,\mathcal S}{\sqrt{\ell^{2}+1}}\right)+f_{11}^{AdS_{4}} (\ell)+\frac{c^{\rm AdS_{4}}}{\log\left(\frac{8\,\pi\,\mathcal S}{\sqrt{\ell^{2}+1}}\right)}+\cdots ,
\end{equation}
with 
\begin{eqnarray}
&& \hspace{-8mm}f_{10}^{AdS_{4}}(\ell) = \frac{1}{2}\,f_{10}^{AdS_{5}}(\ell), \quad 
f_{11}^{AdS_{4}}(\ell) = \frac{1}{2}\,f_{11}^{AdS_{5}}(\ell), \quad
 c^{AdS_{4}}(\ell) = 2\,c^{AdS_{5}}(\ell) = -\frac{\pi}{6\,(\ell^{2}+1)}.
\end{eqnarray}
Notice that the simple $\frac{1}{2}$ rule for the leading two terms is in agreement with the result of \cite{Beccaria:2009wb}. The correction $\sim 1/\log \mathcal S$ comes from the anomaly terms. It is twice bigger than in SYM. For a clear explanation of this fact see \cite{folded}.

\section{Relation with marginality condition}
\label{8}

Let us define $\Lambda\equiv\lambda$ in \ads, and $\Lambda = 16\,\pi^{2}\,g^{2}$ in \adscp. The role of $\Lambda$
is to  emphasize the close analogy between the expressions in the two cases.
For the folded string in \ads, the energy admits the following expansion 
\begin{eqnarray}
\label{eq:marginality} 
E^{2} &=& J^{2}+\left(A_{1}\,\sqrt\Lambda+A_{2}+\frac{A_{3}}{\sqrt\Lambda}
+\cdots\right)\,S + \left(B_{1}+\frac{B_{2}}{\sqrt\Lambda}+ \frac{B_{3}}{\Lambda}+\cdots\right)\,S^{2}
+\cdots , \nonumber
\end{eqnarray}
where the following exact formula  for the constants $A_{i}$ has been conjectured in \cite{Basso:2011rs}:
\begin{equation}
A_{1}\,\sqrt\Lambda+A_{2}+\frac{A_{3}}{\sqrt\Lambda}+\cdots = 2\,\sqrt\Lambda\,Y_{J}(\sqrt\Lambda), \qquad
Y_{J}(x) = \frac{d}{dx}\,\log I_{J}(x).
\end{equation}
It is also know that $B_{1} = \frac{3}{2}$ and $B_{2} = \frac{3}{8}-3\,\zeta(3)$  \cite{Gromov:2011bz} .

The expansion (\ref{eq:marginality}) is very convenient since all powers of $S$ have a coefficient with an expansion 
at large $\Lambda$ starting with a more and more suppressed term. The simplicity of (\ref{eq:marginality}) is a special
feature of the folded string with two cusps. If winding is allowed, it is known that such structure is lost
in \cite{Gromov:2011bz} (see also \cite{Beccaria:2012tu,io}).

For the folded string in \adscp, the expansion with fixed $\mathcal J$ has the general form \cite{Gromov:2011bz} $E = \sqrt\Lambda \,\,\,\mathcal E_{0} + \sum_{\ell=0}^{\infty}\frac{1}{(\sqrt\Lambda)^{\ell}}\,
\sum_{p=1}^{\infty}\sum_{q=-2p}^{\infty} v_{pq}^{(\ell)}\,\mathcal J^{q}\,\mathcal S^{p}$,
where the classical energy $\mathcal E_{0}$ is given in (\ref{eq:classical})
and the semiclassical computation provides $v^{(0)}_{pq}$ according to the results in (\ref{eq:GV}).

Expanding $E^{2}$, we find that  (\ref{eq:marginality}) takes  the following form 
\begin{eqnarray}
\lefteqn{E^{2} - J^{2} = } && \nonumber \\
&&+\bigg[
\left(2-\frac{1}{J}\right) \sqrt{\Lambda }+\left(\frac{2 v^{\text{(1)}}_{1,-2}}{J}-1+2 \log
   (2)\right)+\sqrt{\frac{1}{\Lambda }} \left(
   \frac{2 v^{(2)}_{1,-2}}{J}+
   2 v^{\text{(1)}}_{1,-1}+J^2+\frac{J}{2}\right)+
   \cdots
\bigg]\,S+\nonumber \\
&& + \bigg[
\left(\frac{1}{4 J^4}+\frac{1}{2 J^3}\right) \Lambda +\sqrt{\Lambda }
   \left(-\frac{v^{\text{(1)}}_{1,-2}}{J^4}+\frac{2 v^{\text{(1)}}_{1,-2}}{J^3}+\frac{2
   v^{\text{(1)}}_{2,-4}}{J^3}+\frac{1}{2 J^3}-\frac{\log
   (2)}{J^3}\right)+\cdots
\bigg]\,S^{2}
+\cdots . 
\end{eqnarray}
This structure is different from (\ref{eq:marginality}) since higher powers of $S$ are not associated with terms that are 
more and more suppressed at large $\Lambda$.
This is possible since the new terms not present in (\ref{eq:marginality}) are associated with suitable inverse powers of $J$.
The same phenomenon happens for the winded folded string in $AdS_5 \times S^5$ \cite{Gromov:2011bz}. 
Here wrapping corrections are responsible for these terms.

\subsection{Prediction for short states}

We can provide a prediction for the strong coupling expansion of the energy of short states that in principle
could be tested by TBA calculations. To this aim, we 
can start from  our results at fixed $\rho = \mathcal J/\sqrt\mathcal S$, and 
re-expand at large $\Lambda$ the sum of the (scaled) classical energy $\mathcal E_{0}$
and the one-loop contribution (\ref{eq:our-expansion}). The result is 
\begin{equation}
\label{eq:short}
E = (4\,\pi\,g)^{1/2}\,\sqrt{2\,S}-\frac{1}{2}+\frac{\sqrt{2\,S}}{(4\,\pi\,g)^{1/2}}\,\left(
\frac{J\,(J+1)}{4\,S}+\frac{3\,S}{8}-\frac{1}{4}+\frac{1}{2}\,\log(2)\right)+\cdots.
\end{equation}

The same expansion can be written in terms of the redefined coupling $g \longrightarrow g - \frac{\log 2}{4 \pi}$ \cite{McLoughlin:2008he,Abbott:2010yb}
The result is again eq.(\ref{eq:short}) without the $\log(2)$ term.


\section*{References}
\begin{thebibliography}{9}

\bibitem{Gromov:2009tv}
N.~Gromov, V.~Kazakov, and P.~Vieira, {\em Phys.Rev.Lett.} {\bf 103} (2009) 131601, [{\tt arXiv:0901.3753}].

\bibitem{Arutyunov:2009zu}
G.~Arutyunov and S.~Frolov, {\em JHEP} {\bf 0903} (2009) 152, [{\tt arXiv:0901.1417}].

\bibitem{Bombardelli:2009ns}
D.~Bombardelli, D.~Fioravanti, and R.~Tateo, {\em J.Phys.A} {\bf A42} (2009) 375401, [{\tt arXiv:0902.3930}].

\bibitem{Gromov:2009bc}
N.~Gromov, V.~Kazakov, A.~Kozak, and P.~Vieira, {\em Lett.Math.Phys.} {\bf 91} (2010) 265--287, [{\tt arXiv:0902.4458}].

\bibitem{Arutyunov:2009ur}
G.~Arutyunov and S.~Frolov, {\em JHEP} {\bf 0905} (2009) 068, [{\tt arXiv:0903.0141}].

\bibitem{Gromov:2011de}
N.~Gromov, D.~Serban, I.~Shenderovich, and D.~Volin, {\em JHEP} {\bf 1108} (2011) 046, [{\tt arXiv:1102.1040}].

\bibitem{Roiban:2011fe}
R.~Roiban and A.~Tseytlin, {\em Nucl.Phys.} {\bf B848} (2011) 251--267, [{\tt arXiv:1102.1209}].

\bibitem{Vallilo:2011fj}
B.~C. Vallilo and L.~Mazzucato, {\em JHEP} {\bf 1112} (2011) 029, [{\tt arXiv:1102.1219}].

\bibitem{Basso:2011rs}
B.~Basso, [{\tt arXiv:1109.3154}].

\bibitem{Gromov:2011bz} 
  N.~Gromov and S.~Valatka, JHEP {\bf 1203}, 058 (2012) [{\tt arXiv:1109.6305}].

\bibitem{Frolov:2002av}
S.~Frolov and A.~A. Tseytlin, {\em JHEP} {\bf 0206} (2002) 007, [{\tt hep-th/0204226}].

\bibitem{Aharony:2008ug}
O.~Aharony, O.~Bergman, D.~L. Jafferis, and J.~Maldacena, {\em JHEP} {\bf 0810} (2008) 091, [{\tt arXiv:0806.1218}].

\bibitem{Klose:2010ki} 
  T.~Klose, Lett.\ Math.\ Phys.\  {\bf 99}, 401 (2012) [{\tt arXiv:1012.3999}].
  
\bibitem{McLoughlin:2008ms}
T.~McLoughlin and R.~Roiban, {\em JHEP} {\bf 0812} (2008) 101, [{\tt arXiv:0807.3965}].

\bibitem{Alday:2008ut}
L.~F. Alday, G.~Arutyunov, and D.~Bykov, {\em JHEP} {\bf 0811} (2008) 089, [{\tt arXiv:0807.4400}].

\bibitem{Krishnan:2008zs}
C.~Krishnan, {\em JHEP} {\bf 0809} (2008) 092, [{\tt arXiv:0807.4561}].

\bibitem{Gromov:2008fy}
N.~Gromov and V.~Mikhaylov, {\em JHEP} {\bf 0904} (2009) 083, [{\tt arXiv:0807.4897}].

\bibitem{LevkovichMaslyuk:2011ty}
F.~Levkovich-Maslyuk, JHEP {\bf 1205}, 142 (2012) [{\tt arXiv:1110.5869}].

\bibitem{Gromov:2008bz}
N.~Gromov and P.~Vieira, {\em JHEP} {\bf 0902} (2009) 040, [{\tt arXiv:0807.0437}].

\bibitem{Gromov:2008ec}
N.~Gromov, S.~Schafer-Nameki, and P.~Vieira, {\em JHEP} {\bf 0812} (2008) 013, [{\tt arXiv:0807.4752}].

\bibitem{McLoughlin:2008he}
T.~McLoughlin, R.~Roiban, and A.~A. Tseytlin, {\em JHEP} {\bf 0811} (2008) 069, [{\tt arXiv:0809.4038}].

\bibitem{Abbott:2010yb}
M.~C. Abbott, I.~Aniceto, and D.~Bombardelli, {\em JHEP} {\bf 1012} (2010) 040, [{\tt arXiv:1006.2174}].

\bibitem{BassoSlope} 
  B.~Basso,
  [{\tt arXiv:1205.0054}].

\bibitem{GromovSlope} 
  N.~Gromov,
  [{\tt arXiv:1205.0018}].
  
\bibitem{Tseytlin:2003ac}
A.~A. Tseytlin, {\em Nucl.Phys.} {\bf B664} (2003) 247--275, [{\tt hep-th/0304139}].

\bibitem{Beccaria:2012tu}
M.~Beccaria and G.~Macorini, JHEP {\bf 1202}, 092 (2012) [{\tt arXiv:1201.0608}].

\bibitem{Beccaria:2009wb}
M.~Beccaria and G.~Macorini, {\em JHEP} {\bf 0909} (2009) 017, [{\tt arXiv:0905.1030}].

\bibitem{io} 
  M.~Beccaria, C.~A.~Ratti and A.~A.~Tseytlin, J.\ Phys.\ A A {\bf 45}, 155401 (2012) [{\tt hep-th/1201.5033}].

\bibitem{folded} 
  M.~Beccaria, G.~Macorini, C.~Ratti and S.~Valatka,
  JHEP {\bf 1205}, 030 (2012)
  [Erratum-ibid.\  {\bf 1205}, 137 (2012)]
  [{\tt arXiv:1203.3852}].

\end{thebibliography}
\end{document}